# Twelve shear surface waves guided by clamped/free boundaries in magneto-electro-elastic materials

Arman Melkumyan *

*Department of Mechanics, Yerevan State University, Alex Manoogyan Str. 1,*

*Yerevan 375025, Armenia*

**Abstract**

It is shown that surface waves with twelve different velocities in the cases of different magneto-electrical boundary conditions can be guided by the interface of two identical magneto-electro-elastic half-spaces. The plane boundary of one of the half-spaces is clamped while the plane boundary of the other one is free of stresses. The 12 velocities of propagation of these surface waves are obtained is explicit forms. It is shown that the number of different surface wave velocities decreases from 12 to 2 if the magneto-electro-elastic material is changed to a piezoelectric material.

Bleustein (1968) and Gulyaev (1969) and later Danicki (1994) have shown that an elastic shear surface wave can be guided by the free surface and by an embedded conducting plane in piezoelectric materials in class 6 mm. In this paper the existence of pure shear acoustic surface waves guided by the interface of two identical transversely isotropic magneto-electro-elastic (Nan, 1994, Srinivas et al., 2000, Mori and Wuttig, 2002, Yang et al., 2006) half-spaces in class 6 mm is investigated. The plane boundary of the half-space $x_2 < 0$ is clamped while the plane boundary of the half-space $x_2 > 0$ is free of stresses. Discussing different magneto-electrical boundary conditions in the plane boundaries of the half-spaces pure shear surface waves with 12 different velocities of propagation are obtained. It is expected that these waves will have numerous applications in surface acoustic wave devices.

* Tel. (+374 91) 482977.
*E-mail address:* melk_arman@yahoo.com

Let $x_1$, $x_2$, $x_3$ denote rectangular Cartesian coordinates with $x_3$ oriented in the direction of the sixfold axis of a magneto-electro-elastic material in class 6 mm. Introducing electric potential $\varphi$ and magnetic potential $\phi$, so that $E_1 = -\varphi_{,1}$, $E_2 = -\varphi_{,2}$, $H_1 = -\phi_{,1}$, $H_2 = -\phi_{,2}$, the five partial differential equations which govern the mechanical displacements $u_1$, $u_2$, $u_3$, and the potentials $\varphi$, $\phi$, reduce to two sets of equations when motions independent of the $x_3$ coordinate are considered. The equations of interest in the present paper are those governing the $u_3$ component of the displacement and the potentials $\varphi$, $\phi$, and can be written in the form

$$\begin{aligned} &c_{44}\nabla^2 u_3 + e_{15}\nabla^2\varphi + q_{15}\nabla^2\phi = \rho\ddot{u}_3 \ , \\ &e_{15}\nabla^2 u_3 - \varepsilon_{11}\nabla^2\varphi - d_{11}\nabla^2\phi = 0 \ , \\ &q_{15}\nabla^2 u_3 - d_{11}\nabla^2\varphi - \mu_{11}\nabla^2\phi = 0 \ , \end{aligned} \tag{1}$$

where $\nabla^2$ is the two-dimensional Laplacian operator, $\nabla^2 = \partial^2/\partial x_1^2 + \partial^2/\partial x_2^2$, $\rho$ is the mass density, $c_{44}$, $e_{15}$, $\varepsilon_{11}$, $q_{15}$, $d_{11}$ and $\mu_{11}$ are elastic, piezoelectric, dielectric, piezomagnetic, electromagnetic and magnetic constants, and the superposed dot indicates differentiation with respect to time. The constitutive equations which relate the stresses $T_{ij}$ ($i, j = 1,2,3$), the electric displacements $D_i$ ($i = 1,2,3$) and the magnetic induction $B_i$ ($i = 1,2,3$) to $u_3$, $\varphi$ and $\phi$ are

$$\begin{aligned} &T_1 = T_2 = T_3 = T_{12} = 0 \, , \ D_3 = 0 \, , \ B_3 = 0 \ , \\ &T_{23} = c_{44}u_{3,2} + e_{15}\varphi_{,2} + q_{15}\phi_{,2} \ , \quad T_{13} = c_{44}u_{3,1} + e_{15}\varphi_{,1} + q_{15}\phi_{,1} \ , \\ &D_1 = e_{15}u_{3,1} - \varepsilon_{11}\varphi_{,1} - d_{11}\phi_{,1} \ , \quad D_2 = e_{15}u_{3,2} - \varepsilon_{11}\varphi_{,2} - d_{11}\phi_{,2} \ , \\ &B_1 = q_{15}u_{3,1} - d_{11}\varphi_{,1} - \mu_{11}\phi_{,1} \ , \quad B_2 = q_{15}u_{3,2} - d_{11}\varphi_{,2} - \mu_{11}\phi_{,2} \ . \end{aligned} \tag{2}$$

Solving Eqs. (1) for $\nabla^2 u_3$, $\nabla^2\varphi$ and $\nabla^2\phi$ we find that after defining functions $\psi$ and $\chi$ by

$$\psi = \varphi - m u_3 \ , \quad \chi = \phi - n u_3 \ , \tag{3}$$

the solution of Eqs. (1) is reduced to the solution of

$$\nabla^2 u_3 = \rho\tilde{c}_{44}^{-1}\ddot{u}_3 \ , \quad \nabla^2\psi = 0 \ , \quad \nabla^2\chi = 0 \ , \tag{4}$$

where

$$m = \frac{e_{15}\mu_{11} - q_{15}d_{11}}{\varepsilon_{11}\mu_{11} - d_{11}^2}\ , \quad n = \frac{q_{15}\varepsilon_{11} - e_{15}d_{11}}{\varepsilon_{11}\mu_{11} - d_{11}^2}\ , \tag{5}$$

and

$$\begin{aligned} \tilde{c}_{44} &= c_{44} + \left(e_{15}^2\mu_{11} - 2e_{15}q_{15}d_{11} + q_{15}^2\varepsilon_{11}\right)\Big/\left(\varepsilon_{11}\mu_{11} - d_{11}^2\right) \\ &= \overline{c}_{44}^e + \varepsilon_{11}^{-1}\left(d_{11}e_{15} - q_{15}\varepsilon_{11}\right)^2\Big/\left(\varepsilon_{11}\mu_{11} - d_{11}^2\right) \\ &= \overline{c}_{44}^m + \mu_{11}^{-1}\left(d_{11}q_{15} - e_{15}\mu_{11}\right)^2\Big/\left(\varepsilon_{11}\mu_{11} - d_{11}^2\right). \end{aligned} \tag{6}$$

In Eqs. (6) $\tilde{c}_{44}$ is magneto-electro-elastically stiffened elastic constant, $\overline{c}_{44}^e = c_{44} + e_{15}^2/\varepsilon_{11}$ is electro-elastically stiffened elastic constant and $\overline{c}_{44}^m = c_{44} + q_{15}^2/\mu_{11}$ is magneto-elastically stiffened elastic constant. With the analogy to the electro-mechanical coupling coefficient $k_e^2 = e_{15}^2/\left(\varepsilon_{11}\overline{c}_{44}^e\right)$ and the magneto-mechanical coupling coefficient $k_m^2 = q_{15}^2/\left(\mu_{11}\overline{c}_{44}^m\right)$ introduce magneto-electro-mechanical coupling coefficient

$$\begin{aligned} k_{em}^2 &= \tilde{c}_{44}^{-1}\left(e_{15}^2\mu_{11} - 2e_{15}q_{15}d_{11} + q_{15}^2\varepsilon_{11}\right)\Big/\left(\varepsilon_{11}\mu_{11} - d_{11}^2\right) \\ &= e_{15}^2/\left(\varepsilon_{11}\tilde{c}_{44}\right) + \tilde{c}_{44}^{-1}\varepsilon_{11}^{-1}\left(q_{15}\varepsilon_{11} - e_{15}d_{11}\right)^2\Big/\left(\varepsilon_{11}\mu_{11} - d_{11}^2\right) \\ &= q_{15}^2/\left(\mu_{11}\tilde{c}_{44}\right) + \tilde{c}_{44}^{-1}\mu_{11}^{-1}\left(e_{15}\mu_{11} - q_{15}d_{11}\right)^2\Big/\left(\varepsilon_{11}\mu_{11} - d_{11}^2\right). \end{aligned} \tag{7}$$

From Eqs. (5)-(7) it follows that

$$\begin{aligned} &e_{15}m + q_{15}n = \tilde{c}_{44}k_{em}^2\ , \quad \varepsilon_{11}m + d_{11}n = e_{15}\ , \quad d_{11}m + \mu_{11}n = q_{15}\ , \\ &\left(e_{15}\mu_{11} - q_{15}d_{11}\right)m = \tilde{c}_{44}\mu_{11}k_{em}^2 - q_{15}^2\ , \\ &\left(q_{15}\varepsilon_{11} - e_{15}d_{11}\right)n = \tilde{c}_{44}\varepsilon_{11}k_{em}^2 - e_{15}^2\ . \end{aligned} \tag{8}$$

Using the introduced functions $\psi$ and $\chi$ and the magneto-electro-elastically stiffened elastic constant, the constitutive Eqs. (2) can be written in the following form:

$$\begin{aligned} &T_{23} = \tilde{c}_{44}u_{3,2} + e_{15}\psi_{,2} + q_{15}\chi_{,2}\ , \quad T_{13} = \tilde{c}_{44}u_{3,1} + e_{15}\psi_{,1} + q_{15}\chi_{,1}\ , \\ &D_1 = -\varepsilon_{11}\psi_{,1} - d_{11}\chi_{,1}\ , \quad D_2 = -\varepsilon_{11}\psi_{,2} - d_{11}\chi_{,2}\ , \\ &B_1 = -d_{11}\psi_{,1} - \mu_{11}\chi_{,1}\ , \quad B_2 = -d_{11}\psi_{,2} - \mu_{11}\chi_{,2}\ . \end{aligned} \tag{9}$$

Introduce short notations $e = e_{15}$, $\mu = \mu_{11}$, $d = d_{11}$, $\varepsilon = \varepsilon_{11}$, $q = q_{15}$, $c = c_{44}$, $\overline{c}^e = \overline{c}_{44}^e$, $\overline{c}^m = \overline{c}_{44}^m$, $\tilde{c} = \tilde{c}_{44}$, $w = u_3$, $T = T_{23}$, $D = D_2$, $B = B_2$ and use subscripts $A$ and $B$ to refer to the half-spaces $x_2 > 0$ and $x_2 < 0$, respectively. As the materials in the half-spaces $x_2 > 0$ and $x_2 < 0$ are identical, one has that $e_A = e_B = e$, $\mu_A = \mu_B = \mu$, $d_A = d_B = d$, $\varepsilon_A = \varepsilon_B = \varepsilon$, $q_A = q_B = q$, $c_A = c_B = c$, $\overline{c}_A^e = \overline{c}_B^e = \overline{c}^e$, $\overline{c}_A^m = \overline{c}_B^m = \overline{c}^m$, $\tilde{c}_A = \tilde{c}_B = \tilde{c}$.

The conditions at infinity require that

$$w_A,\ \varphi_A,\ \phi_A \to 0 \ \text{as}\ x_2 \to \infty,$$

$$w_B,\ \varphi_B,\ \phi_B \to 0 \ \text{as}\ x_2 \to -\infty, \tag{10}$$

and the mechanical boundary conditions on the plane boundaries of the half-spaces require that

$$T_A = 0,\ w_B = 0 \ \text{on}\ x_2 = 0. \tag{11}$$

Consider the possibility of a solution of Eqs. (3)-(4) of the form

$$w_A = w_{0A} \exp(-\xi_2 x_2) \exp[i(\xi_1 x_1 - \omega t)],$$

$$\psi_A = \psi_{0A} \exp(-\xi_1 x_2) \exp[i(\xi_1 x_1 - \omega t)], \tag{12}$$

$$\chi_A = \chi_{0A} \exp(-\xi_1 x_2) \exp[i(\xi_1 x_1 - \omega t)],$$

in the half-space $x_2 > 0$ and of the form

$$w_B = w_{0B} \exp(\xi_2 x_2) \exp[i(\xi_1 x_1 - \omega t)],$$

$$\psi_B = \psi_{0B} \exp(\xi_1 x_2) \exp[i(\xi_1 x_1 - \omega t)], \tag{13}$$

$$\chi_B = \chi_{0B} \exp(\xi_1 x_2) \exp[i(\xi_1 x_1 - \omega t)],$$

in the half-space $x_2 < 0$. These expressions satisfy the conditions (10) if $\xi_1 > 0$ and $\xi_2 > 0$; the second and the third of Eqs. (4) are identically satisfied and the first of Eqs. (4) requires

$$\tilde{c}(\xi_1^2 - \xi_2^2) = \rho\omega^2. \tag{14}$$

Now the mechanical boundary conditions (11) together with different magneto-electrical contact conditions on $x_2 = 0$ must be satisfied. The following cases of magneto-electrical contact conditions on $x_2 = 0$ are of our interest in the present paper:

1a) $D_A = D_B = 0$, $\phi_A = \phi_B = 0$;

1b) $D_A = 0$, $\varphi_B = 0$, $\phi_A = \phi_B = 0$;

1c) $D_A = D_B = 0$, $B_B = 0$, $\phi_A = 0$;

1d) $D_A = 0$, $\varphi_B = 0$, $B_B = 0$, $\phi_A = 0$;

2a) $\varphi_A = 0$, $D_B = 0$, $B_A = B_B = 0$;

2b) $\varphi_A = \varphi_B = 0$, $B_A = B_B = 0$;

2c) $\varphi_A = 0$, $D_B = 0$, $B_A = 0$, $\phi_B = 0$;

2d) $\varphi_A = \varphi_B = 0$, $B_A = 0$, $\phi_B = 0$;

3a) $\varphi_A = \varphi_B = 0$, $\phi_A = \phi_B = 0$;

3b) $\varphi_A = 0$, $D_B = 0$, $\phi_A = \phi_B = 0$;

3c) $\varphi_A = \varphi_B = 0$, $B_B = 0$, $\phi_A = 0$;

3d) $\varphi_A = 0$, $D_B = 0$, $B_B = 0$, $\phi_A = 0$;

4) $D_A = D_B = 0$, $B_A = B_B$, $\phi_A = \phi_B$; (15)

5) $D_A = D_B$, $\varphi_A = \varphi_B$, $B_A = B_B = 0$;

6) $D_A = D_B$, $\varphi_A = \varphi_B$, $B_A = B_B$, $\phi_A = \phi_B$;

7) $D_A = D_B$, $\varphi_A = \varphi_B$, $\phi_A = \phi_B = 0$;

8) $\varphi_A = \varphi_B = 0$, $B_A = B_B$, $\phi_A = \phi_B$;

9) $D_A = 0$, $\varphi_B = 0$, $B_A = B_B$, $\phi_A = \phi_B$;

10) $D_A = D_B$, $\varphi_A = \varphi_B$, $B_A = 0$, $\phi_B = 0$;

11) $D_A = D_B$, $\varphi_A = \varphi_B$, $B_B = 0$, $\phi_A = 0$;

12) $\varphi_A = 0$, $D_B = 0$, $B_A = B_B$, $\phi_A = \phi_B$;

13a) $D_A = D_B = 0$, $B_A = B_B = 0$;

13b) $D_A = 0$, $\varphi_B = 0$, $B_A = B_B = 0$;

13c) $D_A = D_B = 0$, $B_A = 0$, $\phi_B = 0$;

13d) $D_A = 0$, $\varphi_B = 0$, $B_A = 0$, $\phi_B = 0$.

Each of the 25 groups of conditions in Eqs. (15) together with Eqs. (11), Eqs. (12)-(13) leads to a system of six homogeneous algebraic equations for $w_{0A}$, $\psi_{0A}$, $\chi_{0A}$, $w_{0B}$, $\psi_{0B}$, $\chi_{0B}$, the existence of nonzero solution of which requires that the determinant of that system must be equal to zero. This condition for the determinant together with Eq. (14) determines the surface wave velocities $V_s = \omega/\xi_1$. In the case of 1a) of Eqs. (15) this procedure leads to a surface wave with velocity

$$V_{s1}^2 = (\tilde{c}/\rho)\left(1 - \left[k_{em}^2 - e^2/(\tilde{c}\varepsilon)\right]^2\right). \tag{16}$$

The same velocity is obtained in the cases 1b), 1c) and 1d). Each of the cases 2a), 2b), 2c) and 2d) leads to a surface wave with velocity

$$V_{s2}^2 = (\tilde{c}/\rho)\left(1 - \left[k_{em}^2 - q^2/(\tilde{c}\mu)\right]^2\right), \tag{17}$$

and each of the cases 3a), 3b), 3c) and 3d) leads to a surface wave with velocity

$$V_{s3}^2 = (\tilde{c}/\rho)\left(1 - k_{em}^4\right). \tag{18}$$

Each of the cases 4) to 12) leads to its own one surface wave and the corresponding surface wave velocities are

$$V_{s4}^2 = (\tilde{c}/\rho)\left(1 - \tfrac{1}{4}\left[k_{em}^2 - e^2/(\tilde{c}\varepsilon)\right]^2\right);$$

$$V_{s5}^2 = (\tilde{c}/\rho)\left(1 - \tfrac{1}{4}\left[k_{em}^2 - q^2/(\tilde{c}\mu)\right]^2\right);$$

$$V_{s6}^2 = (\tilde{c}/\rho)\left(1 - \tfrac{1}{4}k_{em}^4\right);$$

$$V_{s7}^2 = (\tilde{c}/\rho)\left(1 - \left[k_{em}^2 - \tfrac{1}{2}e^2/(\tilde{c}\varepsilon)\right]^2\right);$$

$$V_{s8}^2 = (\tilde{c}/\rho)\left(1 - \left[k_{em}^2 - \tfrac{1}{2}q^2/(\tilde{c}\mu)\right]^2\right); \tag{19}$$

$$V_{s9}^2 = (\tilde{c}/\rho)\left(1 - \left(\frac{\varepsilon\mu}{2\varepsilon\mu - d^2}\right)^2\left[k_{em}^2 - e^2/(\tilde{c}\varepsilon)\right]^2\right);$$

$$V_{s10}^2 = (\tilde{c}/\rho)\left(1 - \left(\frac{\varepsilon\mu}{2\varepsilon\mu - d^2}\right)^2\left[k_{em}^2 - q^2/(\tilde{c}\mu)\right]^2\right);$$

$$V_{s11}^2 = (\tilde{c}/\rho)\left(1 - \left[k_{em}^2 - \frac{\varepsilon\mu}{2\varepsilon\mu - d^2}e^2/(\tilde{c}\varepsilon)\right]^2\right);$$

$$V_{s12}^2 = (\tilde{c}/\rho)\left(1 - \left[k_{em}^2 - \frac{\varepsilon\mu}{2\varepsilon\mu - d^2}q^2/(\tilde{c}\mu)\right]^2\right).$$

The cases 13a) to 13d) do not lead to any surface wave.

If the magneto-electro-elastic material degenerates to a piezoelectric material, so that $q \to 0$, $d \to 0$, the surface waves that have velocities $V_{s1}$, $V_{s4}$, $V_{s9}$ disappear and

$$V_{s2},\ V_{s3},\ V_{s8},\ V_{s12} \to V_{bg} = \sqrt{(\bar{c}^e/\rho)\left(1 - k_e^4\right)};$$

$$V_{s5}\,,\ V_{s6}\,,\ V_{s7}\,,\ V_{s10}\,,\ V_{s11} \to \sqrt{\left(\bar{c}^e/\rho\right)\left(1-\tfrac{1}{4}k_e^4\right)}\,, \qquad (20)$$

so that the number of different surface wave velocities decreases from 12 to 2 when the magneto-electro-elastic material is changed to a piezoelectric material. In Eq. (20) $V_{bg}$ is the Bleustein-Gulyaev surface wave speed.